\documentclass[apj]{emulateapj}

\shorttitle{Radial color gradients of SDSS Early-type galaxies}
\shortauthors{Suh et al.}

\begin{document}

\title{Demography of SDSS Early-type galaxies \\
        from the perspective of radial color gradients}
        
\author{Hyewon Suh, Hyunjin Jeong, Kyuseok Oh and Sukyoung K. Yi}
\affil{Department of Astronomy, Yonsei University, Seoul 120-749, Korea}

\and

\author{Ignacio Ferreras}
\affil{Mullard Space Science Laboratory, University College London, \\
      Holmbury St Mary, Dorking, Surrey RH5 6NT, UK}

\and

\author{Kevin Schawinski\altaffilmark{$\dagger$}}
\affil{Yale Center for Astronomy and Astrophysics, Yale University, \\
         P.O. Box 208121, New Haven, CT 06520, U.S.A.}

\altaffiltext{$\dagger$}{Einstein Fellow}

\begin{abstract}
We have investigated the radial $g-r$ color gradients of
early-type galaxies in the Sloan Digital Sky Survey (SDSS) DR6 in the
redshift range $0.00 \le z \le 0.06$. The majority of massive
early-type galaxies show a negative color gradient (red-cored) as
generally expected for early-type galaxies. On the other hand, roughly
30~per cent of the galaxies in this sample show a positive color
gradient (blue-cored). These ``blue-cored'' galaxies often show
strong H$\beta$ absorption line strengths and/or emission line ratios
that are indicative of the presence of young stellar
populations. Combining the optical data with {\it Galaxy Evolution Explorer}
({\it GALEX}) ultraviolet photometry, we find that {\sl all} blue-cored
galaxies show UV$-$optical colors that can only be explained by young
stellar populations. This implies that most of the residual star
formation in early-type galaxies is centrally concentrated. Blue-cored galaxies
are predominantly low velocity dispersion systems, and tend to live in
lower density regions. 
A simple model shows that the observed positive color gradients (blue-cored) 
are visible only for a billion years after a star formation episode for the typical strength
of recent star formation.  The observed effective radius
decreases and the mean surface brightness increases due to this
centrally-concentrated star formation episode. As a result, the majority
of blue-cored galaxies may lie on different regions in the Fundamental Plane from red-cored
ellipticals. However, the position of the blue-cored
galaxies on the Fundamental Plane cannot be solely attributed to
recent star formation but require substantially lower velocity
dispersion. Our results based on the optical data are consistent with
the residual star formation interpretation of Yi and collaborators
which was based on {\it GALEX} UV data.  We conclude that a low-level of
residual star formation persists at the centers of most of low-mass
early-type galaxies, whereas massive ones are mostly quiescent 
systems with metallicity-driven red cores.
\end{abstract}

\keywords{galaxies: elliptical and lenticular, cD -- galaxies:
evolution -- galaxies: photometry -- galaxies: structure}

%%%%%%%%%%%%%%%%%%%%%%%%%%%%%%%%%%%%%%%%%%%%
\section{Introduction}

Deciphering the star formation histories of early-type galaxies is a
key issue to understanding the fundamental processes of galaxy
formation and evolution. Spatially resolved analyses can shed light on
the build up of the stellar populations.  Observations have shown that
the colors of early-type galaxies get bluer from the center outwards
\citep{Boroson83, Franx90, Peletier90a, Peletier90b, Michard99,
Idiart02, Propris05, Wu05, LC09}. Furthermore, the lack of a strong trend in the
gradient over a wide range of lookback time \citep{Ferreras09}
suggests that this negative color gradient originates from
metallicity variations, with the center being more metal rich.
This is often taken as evidence pointing to a simple
evolutionary model for early-type galaxies. This classical
collapse model \citep{ELS, Larson74a, Larson74b, Larson75, Carlberg84}
suggests that early-type galaxies form in highly efficient starbursts
at high redshifts and have evolved without any substantial amount of
subsequent star formation.

While early-type galaxies were traditionally considered to be
dynamically simple stellar systems with homogeneous stellar
populations, it is now clear that they are likely to have undergone
complex and varied formation histories. Recent studies on the color
gradients of early-type galaxies have shown that a significant fraction
feature positive color gradients \citep[i.e. blue cores,][]{Michard99,
Menanteau01, Ferreras05, Elmegreen05}. To explain these positive color gradients,
it is necessary to have age gradients in addition to metallicity
gradients \citep{Silva94, Michard05}. These age gradients could be a
natural result of the hierarchical merger scenario, where early-type
galaxies often form as a result of galaxy mergers
\citep[e.g.][]{tt72}. In this model, early-type galaxies form as the
result of successive mergers and are thought to have continued or
episodic star formation events. Indeed, \citet{Menanteau01a} found
that a remarkably large fraction of early-type galaxies at high
redshift exhibit positive gradients as a result of recent merging
or an inflow of material. \citet{Ferreras05} also reported that
about one-third of field early-type galaxies at $z \sim 0.7$ have
blue cores. These findings suggest that recent episodes of
star formation take place at the center of these galaxies \citep[see
also][]{Menanteau01, Friaca01, Marcum04, Elmegreen05}.

These findings are also related to the positive gradients found in post-starburst galaxies
(e.g., E+A galaxies; \citealt{Dressler83, Dressler92, Pogg99, Norton01, Ba01, GT04, Yama05} ).
\citet{Yang08} suggests that E+A galaxies are remnants of galaxy-galaxy 
interactions/mergers and ultimately evolve into early-type galaxies 
\citep[see also][]{GG72,Zab96,Blake04}. They demonstrated that a 
large fraction of E+A galaxies have positive color gradients 
and these gradients can evolve into negative gradients as the populations age.

The {\it Galaxy Evolution Explorer} ({\it GALEX}) ultraviolet (UV) filters
allow us to detect very small amounts of young stars within an old
stellar population.
Therefore, {\it GALEX} data provide a unique opportunity to study the recent star
formation history of galaxies. Many studies have shown that a
significant fraction of early-type galaxies exhibit enhanced UV light
as a sign of recent star formation \citep{Yi05, Salim07, Donas07,
S07a, Kaviraj07, Kaviraj08}. 
The observed positive color gradients in the center of some early-type
galaxies could be naturally accounted for by these young stellar populations.

The role of environment in the formation and evolution of galaxies is
another key issue. The physical properties of early-type galaxies
correlate with their environment. \citet{Dressler80} pointed out that
the environment has an effect on galaxy morphology, whereby the
abundance of early-type galaxies increases in dense environments.
This morphology-density relation implies that early-type galaxies are
more common in clusters. \citet{Bamford09} recently suggested that the
color-density relation is stronger than the morphology-density
relation at fixed stellar mass. Most low stellar mass galaxies of any
morphology are blue in low-density environments but red in
high-density regions. They also found that there is a substantial
fraction of galaxies in low-density environments with blue colors but
with an early-type morphology.  It is also worth considering the
effect of the nearest neighbor galaxies on galaxy properties.
\citet{PC09} found that late-type neighbors enhance the star formation
activity of galaxies and these effects occur within the virial radius.
Furthermore, on a sample of close pairs only comprising early-type
galaxies, \citet{Rogers09} found an enhancement of residual star formation
with respect to a field sample, and a correlation between (mild) active
galactic nuclei (AGN) activity and pair separation.

The Fundamental Plane (FP) is a key scaling relation for early-type
galaxies. It is a two-dimensional plane in the three-dimensional
manifold spanned by global structural parameters (effective radius,
surface brightness and stellar velocity dispersion; e.g.\
\citealt{detal87,dd87}). Through changes in the surface brightness
distribution, recent star formation can change the position of a
galaxy on the FP. \citet{Choi09} found that the FP of E+A galaxies is
different from that of quiescent early-type galaxies, and this is most
likely due to a recent starburst in the central
regions. \citet{Jeong09} also suggested that a dominant fraction of
the tilt and scatter of the UV FP is due to the presence of young
stars in galaxies with blue colors in the NUV, that are generally low-mass systems.

In this paper, we measure the radial color gradients of early-type
galaxies at $0.00 \le z \le 0.06$ from Data Release 6 of the Sloan
Digital Sky Survey (SDSS, \citealt{York00, Stoughton02, Adelman08}) and
use them to investigate the star formation history in early-type
galaxies. Hereafter, galaxies with a negative color gradient are
labelled as ``red-cored galaxies''.  Similarly, galaxies
with a positive gradient are labelled as ``blue-cored galaxies''.
In \S~\ref{sec:sample}, we describe the sample and data
reduction. In \S~\ref{sec:properties}, we present the properties of
early-type galaxies such as the color-magnitude relation, UV
luminosities, velocity dispersion, H$\beta$ line strengths, emission
line diagnostics, FP, concentration index and
environment with respect to radial color gradients.  In
\S~\ref{sec:model}, we quantify and discuss the recent star formation
in early-type galaxies and revisit the effects of star formation on
the FP. Finally, we summarize our findings in
\S~\ref{sec:summary}.

Throughout this paper we assume a $\Lambda$CDM cosmology with
$\Omega_{m} = 0.3$ and H$_{0} = 70$ km s$^{-1}$ Mpc$^{-1}$.
\\

%%%%%%%%%%%%%%%%%%%%%%%%%%%%%%%%%%%%%%%%%%%%
\section[]{Sample selection and Data reduction}
\label{sec:sample}
%%%%%%%%%%%%%%%%%%%%%%%%%%%%%%%%%%%%%%%%%%%%
\subsection{Early-type Galaxy Selection in SDSS}

%figure 1
\begin{figure}
\centering
\includegraphics[width=0.5\textwidth]{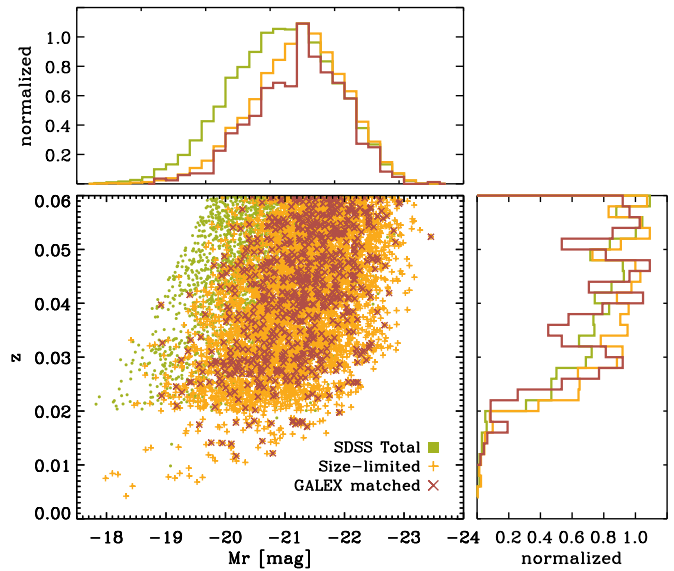}
\caption{Our sample of SDSS DR6 visually classified early-type
galaxies at $0.00 \le z \le 0.06$. Green points in the background
indicate all early-type galaxies. Orange points represent the
size-limited sample (a criterion imposed to obtain robust color
gradients).  We plot galaxies matched with {\it GALEX} as red points.
We also plot histograms for M$_{\rm r}$ and z.}
\label{fig:sample}
\end{figure}

% figure 2
\begin{figure*}
\centering
\includegraphics[width=1\textwidth]{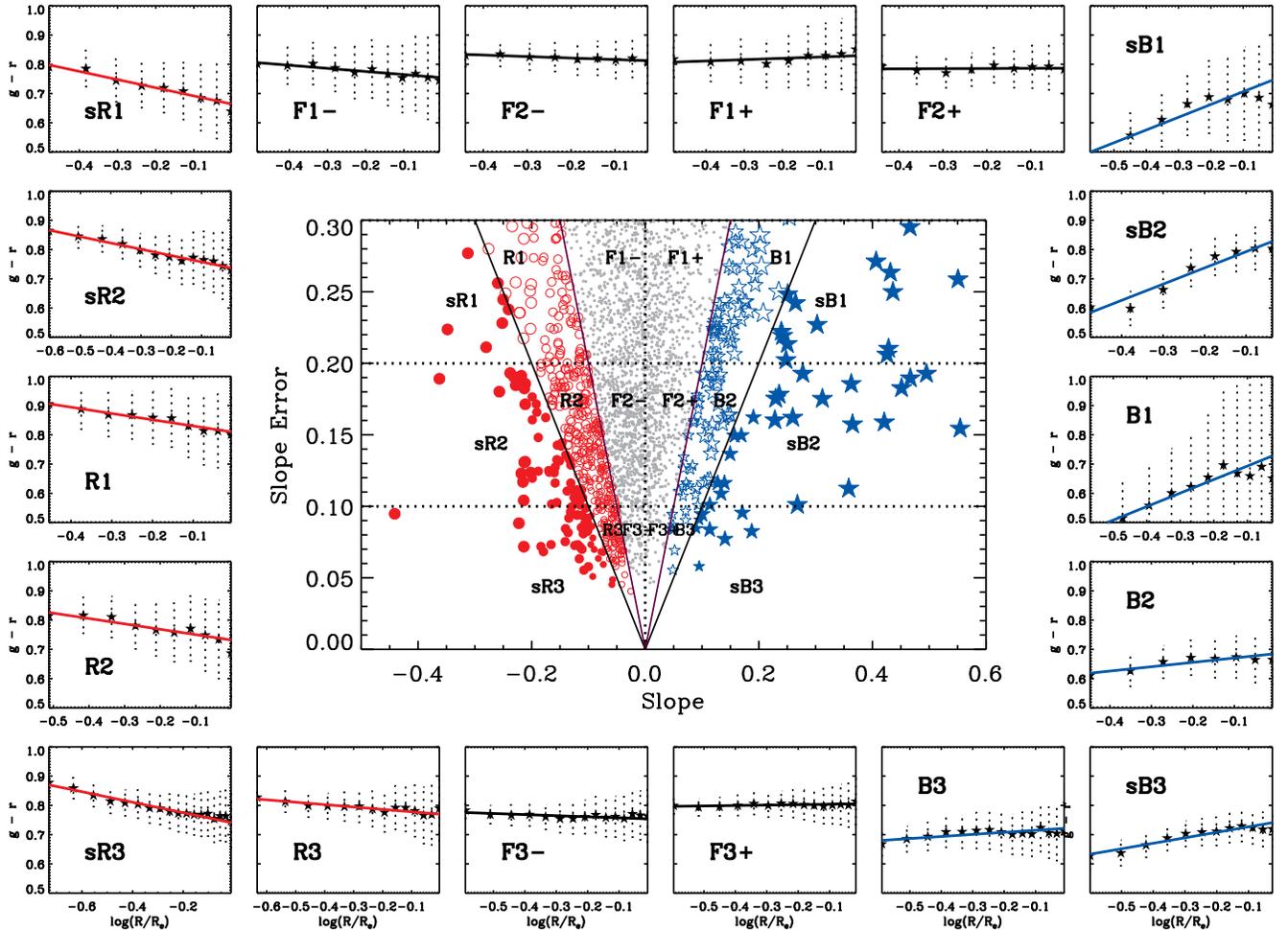}
\caption{Slope errors of early-tye galaxies versus
$g-r$ radial slope -- defined as d($g-r$)/d$\log$(R/R$_{\rm e}$) -- in the middle. 
The black solid lines indicate 0.5$\sigma$ (two sloped solid lines
close to the vertical dotted line) and 1$\sigma$ confidence
level lines, respectively. We select blue-cored galaxies
that have steeper slope than the 0.5$\sigma$ confidence level line
(blue stars). Red circles represent galaxies with negative $g-r$
gradients (red-cored) that have steeper slope than the 0.5$\sigma$ confidence
level. Gradients from 0.5$\sigma$ and 1$\sigma$ level confidence are
plotted as open and filled symbols, respectively. Bigger symbols
represent steeper slopes. Sample plots of radial $g-r$ color
gradients with different slope and slope errors are shown around the
middle panel.}
\label{fig:err_slope}
\end{figure*}

The Sloan Digital Sky Survey (SDSS) targets a large portion of the
northern sky providing photometry in {\it u, g, r, i} and {\it z}
bands. Furthermore, the SDSS spectroscopic survey includes virtually
all extended objects with Petrosian magnitude {\it r} $<$ 17.77 that
escape fiber collision problems \citep{Strauss02}. We start by selecting all
unsaturated galaxies with spectra in the redshift interval $0.00 \le z \le 0.06$
and apply the apparent {\it r} band Petrosian magnitude cut of 17.50
to ensure the imaging quality suffices for the
morphological classification of the sample. The SDSS photometry is
given in the AB system \citep{Oke83}. However, it is known that the
photometric zero points are slightly offset with respect to the AB
standard. Therefore, we use the SDSS photometric flux calibration
algorithm to obtain a calibrated magnitude and apply the offset between
the SDSS system and the AB system \citep{Bohlin01}. We also correct for
Galactic extinction using the \citet{Schlegel98} maps provided by the
SDSS pipeline and apply a k-correction as described in \citet{Blanton07}.
For galaxy luminosities, we used the SDSS
\texttt{petroMag}, which is a good estimate of the total light, while
the colors were derived from SDSS \texttt{modelMags}, which give a  more
reliable measure of unbiased colors regardless of any color gradients
\citep{Strauss02}.

Our next step involves the visual classification of the sample. We
opted for a morphological classification instead of those based on
color or concentration, as it reduces the bias that those methods
have with respect to the presence of young stars.  We make our initial
selection based on the \texttt{fracDev} parameter, which is the weight
of the deVaucouleur's profile in the best composite (deVaucouleur's +
exponential) fit to the image in each band. We extract all galaxies
which have \texttt{fracDev} $\ge$ 0.95 in {\it g, r} and {\it i}
bands, following \citet{Yi05} \citep[see also][] {Kaviraj07, S07a}.
On this sample, we perform visual inspection to exclude galaxies with
spiral arms or distorted features. In addition, we exclude small
galaxies to guarantee a reliable measurement of the radial color
gradients (see \S 2.2).  The final sample comprises a total
of 5,002 early-type galaxies. Table~\ref{tbl:sampling_crit} presents
a summary of the selection criteria.

We cross-match the detections in the {\it GALEX} Medium Imaging Survey (MIS)
with the SDSS early-type sample from 704 fields, with a 4 arcsec
tolerance. In the case of multiple galaxy matches within a given
matching radius, we select the UV source with the smallest angular
separation from the position of the SDSS galaxy. We also de-redden
the colors with respect to Galactic extinction using
A$_{NUV}$~=~8.741~$\times$~E(B$-$V) \citep{Wyder05}
with the reddening maps from \citet{Schlegel98} and apply a
k-correction. In Figure~\ref{fig:sample}, we plot all visually-inspected
early-type galaxies as green points. The orange points correspond to
the final size-limited sample (see \S 2.2 for details).  
Those galaxies in the final sample
with a UV detection from {\it GALEX} are shown as red points.
We also plot histograms for M$_{\rm r}$ and z.

%table1
\begin{table}
 \centering
  \caption{Summary of sampling criteria}
  \begin{tabular}{l l}
  \tableline \tableline
    \multicolumn{1}{c}{Criterion} & \multicolumn{1}{c}{Reason} \\
  \tableline
     r $<$ 17.50   		&  Ensure photometric depth \\
     0.00 $\le$ z $\le$ 0.06  	&  Limit redshift for morphological \\
   				& classification \\
     \texttt{fracDev$_{g,r,i}$} $\ge$ 0.95          & Robust morphological classification  \\
     R$_{e}$ $-$ R$_{\rm PSF}$ $>$ 1.188''         & Guarantee a reliable measurement of \\
     				& radial color gradients (having more \\
     				& than 3 fitting points between \\
     				& the PSF width and effective radius) \\
 \tableline \tableline
\end{tabular} \label{tbl:sampling_crit}
\end{table}

%%%%%%%%%%%%%%%%%%%%%%%%%%%%%%%%%%%%%%%%%%%%
\subsection[]{Measurement of color gradients}

The SDSS provides the corrected frames (flat-fielded,
sky-subtracted, and calibrated sub-images corrected for bad columns
and cosmic rays) in five bands with $2046 \times 1489$ pixels,
sampled with $0.\arcsec396 \times 0.\arcsec396$ pixels. Although the
images are delivered pre-processed, we perform our own sky
subtraction by measuring the sky level using the IDL ``\texttt{sky}'' Routine
which adopts the technique from the DAOPHOT
because the SDSS photometric pipeline tends to overestimate the sky
background \citep{Linden07}. We use the modal values for sky estimation.

We carry out the surface photometry of our sample of early-type galaxies
in {\it g} and {\it r} bands (both of which are redward of the Balmer break)
by measuring the surface brightness along
elliptical annuli, implemented in the \texttt{ELLIPSE} task within the
\texttt{STSDAS ISOPHOTE} package in IRAF\footnote{http://iraf.noao.edu/}
(Image Reduction and Analysis Facility). 
We fixed the center of the isophotes to the center
of the light distribution, and left the position angle (PA),
ellipticity ({\it e}) and surface brightness ($\mu$) free as a
function of radius. Ellipses were fit to the higher S/N {\it r} band image
only, and then overlaid on the {\it g} band images so that the colors
are extracted from the same regions.

% figure 3
\begin{figure}
\centering
\includegraphics[width=0.5\textwidth]{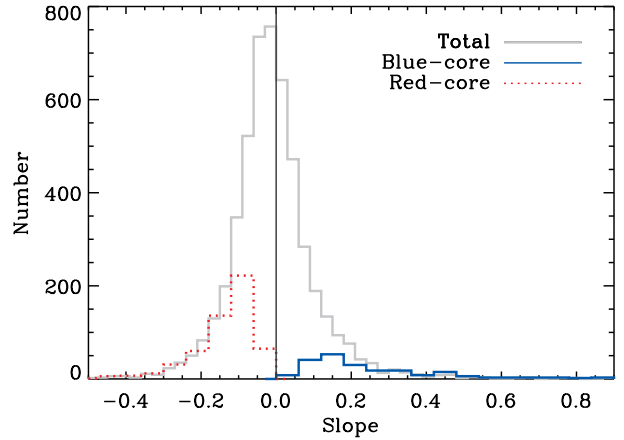}
\caption{The distribution of $g-r$ color gradients.
From the total sample of 5,002 early-type galaxies, we select a subsample of 761 
galaxies with significant color gradients.
The gray histogram represents the entire sample of 5,002 galaxies (most of them
showing a very small color gradient).
The blue solid and red dotted histograms represent blue- and red-cored galaxies,
respectively.} \label{fig:Slopehist}
\end{figure}

%table2
\begin{deluxetable*}{lrrr}
\centering
\tablewidth{0pt}
\tablecaption{$g-r$ color gradients}
\tablehead{
\colhead{Sample} & 
\colhead{mean(slope$^{a}$)} & 
\colhead{median(slope)} &  
\colhead{mean($\Delta$slope)}
}
\startdata
   Total (5,002) & -0.009 & -0.017 & 0.21 \\
   $\Delta$slope$^{b} <$ 0.1 (294) & -0.035 & -0.036 & 0.08 \\
   $\Delta$slope $<$ 0.2 (1,555) & -0.026 & -0.027 & 0.13 \\
   $\Delta$slope $<$ 0.3 (2,564) & -0.019 & -0.023 & 0.18 \\
 \\
 Blue-cored$^{c}$ (216) & 0.25 & 0.18 & 0.24 \\
 Strong blue-cored$^{d}$ (66) & 0.41 & 0.41 & 0.25 \\
 Red-cored$^{e}$ (545) & -0.14 & -0.12 & 0.14 \\
\enddata
\tablenotetext{a}{Slope = $\rm \frac{d(g-r)}{dlog(R/R_{e})}$}
\tablenotetext{b}{Slope errors of our galaxies as shown in Figure~\ref{fig:err_slope}.}
\tablenotetext{c}{Blue-cored galaxies shown in B1, B2, B3, sB1, sB2, sB3 in Figure~\ref{fig:err_slope} }
\tablenotetext{d}{Blue-cored galaxies shown in sB1, sB2, sB3 in Figure~\ref{fig:err_slope}}
\tablenotetext{e}{Red-cored galaxies shown in R1, R2, R3, sR1, sR2, sR3 in Figure~\ref{fig:err_slope}}
\label{tbl:color_grad}
\end{deluxetable*}

From the radial surface brightness profiles, we derive $g-r$ color
gradients with a least-squares fit. The radial range of the fit is
adjusted individually for each galaxy.  The minimum radial coordinate
of the fit is given by the FWHM of the point spread function
(PSF), in order to minimize seeing effects. The PSF widths in the {\it g}
and {\it r} bands are different. We select the worse case between
them. Moreover, we use in the fit pixels out to an effective radius
(R$_{\rm e}$), because the data usually become noisy at larger radii. From
this sample, we select those galaxies with more than 3 points
(i.e. $\sim$1.188 arcsec) within the fitting range considered, to
guarantee a reliable measurement of the radial color gradient --
defined by its slope,
$\equiv$ d($g-r$)/d$\log$(R/R$_{\rm e}$). Additionally, we perform a Monte
Carlo simulation comprising 10$^{3}$ realizations adding noise to each
point compatible with the observations, to assess the accuracy of the
color gradients measured.  We plot the uncertainty of the slope versus
color gradient in the middle of
Figure~\ref{fig:err_slope}.  The size of the symbols is proportional
to the slope. We adopt a 0.5$\sigma$ confidence level (two sloped solid
lines close to the vertical dotted line) to pick out galaxies that 
show statistically-significant color
gradients. Using this criterion, we identify hereafter the 
blue-cored galaxies in Figure~\ref{fig:err_slope} 
as blue stars. We 
label galaxies with slope~$<$~$-$0.5 $\times$ (slope error) as 
red-cored galaxies and show them as red circles thourghout this
paper. Furthermore, open and filled symbols indicate the galaxies
with slopes steeper than 0.5$\sigma$ and 1$\sigma$ confidence
levels, respectively. Table~\ref{tbl:color_grad} presents the mean
values of the slope for different subsamples as shown in
Figure~\ref{fig:err_slope}.  Note that galaxies with smaller
uncertainties usually have negative slopes (red-cored). Furthermore, 
for a significant fraction of the sample the error estimates 
are rather large, a result of the shallow imaging of the SDSS survey.

We plot the distribution of $g-r$ color gradients in
Figure~\ref{fig:Slopehist}. The gray line shows the histogram for a
total of 5,002 galaxies. The blue solid and red dotted histograms correspond to
blue- and red-cored galaxies, respectively. The overall fraction
of blue- and red-cored galaxies within our criterion is
roughly 4 per cent (216/5,002) and 11 per cent (545/5,002),
respectively. We note that roughly 30 per cent of the galaxies with
detectable color gradients have positive values (i.e. blue-cored). 
In other words, more early-type galaxies tend to have 
negative color gradients (i.e. red-cored).

To measure the line strengths, we use the GANDALF\footnote{http://star.herts.ac.uk/~sarzi/}
(Gas AND Absorption Line Fitting) package
\citep{Sarzi06}. It is important to remove the contribution from
emission lines in order to track the true values of absorption from
the underlying stellar populations. Contamination due to emission lines
from extended regions of ionized gas exists in roughly 75 per cent of
early-type galaxies. The GANDALF measurements, such as standard Lick
absorption line indices, emission line strengths and velocity
dispersions of the SDSS galaxies are described in detail in Oh et
al. (in preparation).

%%%%%%%%%%%%%%%%%%%%%%%%%%%%%%%%%%%%%%%%%%%%
\section[]{Results}
\label{sec:properties}
%%%%%%%%%%%%%%%%%%%%%%%%%%%%%%%%%%%%%%%%%%%%
\subsection{Color-magnitude relation}

% figure 4
\begin{figure*}
\centering
\includegraphics[width=0.8\textwidth]{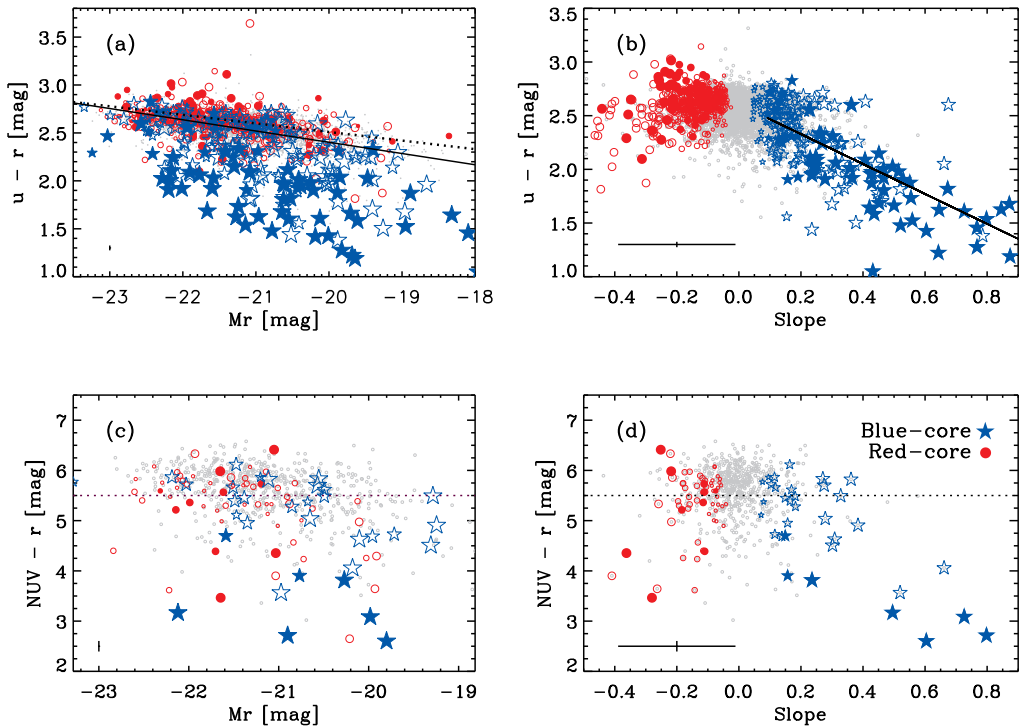}
\caption{Gray dots in the background represent all galaxies.
We plot blue-cored and red-cored galaxies with blue stars and red
circles, respectively. A typical error bar is shown as comparison.
(a) $u-r$ vs. M$_r$ color-magnitude relation.
Black solid and dotted line indicate the least-square fit to the total sample
and only to the red-cored galaxies, respectively.
(b) $u-r$ color vs. $g-r$ radial slope.
The black line indicates the least-square fit of blue-cored galaxies.
(c) NUV$-$r color-magnitude relation.
(d) NUV$-$r color vs. $g-r$ slope.
The symbols are the same as Figure~\ref{fig:err_slope}.}
\label{fig:cmr}
\end{figure*}

The color-magnitude relation is widely used to study the star
formation history of early-type galaxies \citep[see
e.g.][]{ble92b,sed98,fcs99}. The correlation between metallicity and
luminosity is the main reason of this relation \citep{KA97}. The
optical color-magnitude relation of early-type galaxies reveals a small
scatter around the mean relation (e.g. \citealt{ble92, Ellis97,
Dokkum00}). Figure~\ref{fig:cmr}(a) shows the $u-r$
color-magnitude relation. We note that the $u$ band is particularly
sensitive to the presence of young stellar populations. 
The figure also shows the least-squares fit to the
whole sample (solid) and only to the red-cored galaxies (dotted).
The gray points, blue stars and red circles indicate the whole sample,
blue-cored and red-cored galaxies, respectively.  A cursory
inspection of this diagram shows that most blue-cored galaxies are located in
the blue cloud, while red-cored galaxies reside in the red sequence, in
agreement with recent studies of early-type galaxies at moderate
redshift \citep{Ferreras09}. It should also be noted that a
non-negligible fraction of the scatter in the optical color magnitude
relation is due to blue-cored galaxies.

We also show integrated $u-r$ color versus
the slope of the $g-r$ color gradient in Figure~\ref{fig:cmr}(b). 
It reveals that a strong correlation is present
amongst the blue-cored galaxies, with a correlation
coefficient $r=$ 0.76. This implies that whenever star formation is
present in early-type galaxies, it is mostly concentrated in the
central regions, at least within the effective radius. We recall
however that some galaxies show significantly bluer $u-r$ colors even
though they have shallower slopes. These galaxies show that the $g-r$
color is significantly bluer out to at least the effective radius,
suggesting spatially extended star formation. Meanwhile, we note that
some red-cored galaxies with steep slopes show slightly blue $u-r$ colors. In
this case, the presence of {\em off-centered} young stellar populations can
explain both the bluer integrated color and the steep color gradient.

The NUV color-magnitude relation (see, e.g., \citealt{Yi05,
Kaviraj07,S07a}) is a particularly efficient tool for tracking recent
star formation, owing to its high sensitivity to young stellar
populations.  In Figure~\ref{fig:cmr}(c, d), we show the NUV$-$r 
color-magnitude relation and NUV$-$r color versus
slope in the $g-r$ radial gradient. The empirical
threshold at NUV$-$r $<$ 5.5 of \citet{Yi05} (based on NGC\,4552) to
indicate recent star formation is shown as a dashed line. We find
that blue-cored galaxies with steep slopes (filled stars)
reside almost exclusively below NUV$-$r $\sim 5.5$,
supporting our findings that most of blue-cored galaxies have
undergone recent star formation, while red-cored galaxies are
relatively red in the NUV. Interestingly, all blue-cored galaxies
show blue NUV$-$optical colors but the converse is not always true.
We note that there are some red-cored galaxies with NUV$-$r $<$
5.5.  This indicates that some early-type galaxies have steep negative
color gradients because of the presence of young stars in the outer
regions.  For example, NGC 2974 -- classified as an early-type galaxy
in the optical -- shows blue UV$-$optical colors at large radii
\citep{Jeong07}. The UV bright outer ring observed in this galaxy is
related to a bar, demonstrating that recent star formation can be
found in the form of a ring in the outer part of an early-type galaxy.

%%%%%%%%%%%%%%%%%%%%%%%%%%%%%%%%%%%%%%%%%%%%
\subsection{Velocity dispersion}

% figure 5
\begin{figure}
\centering
\includegraphics[width=0.5\textwidth]{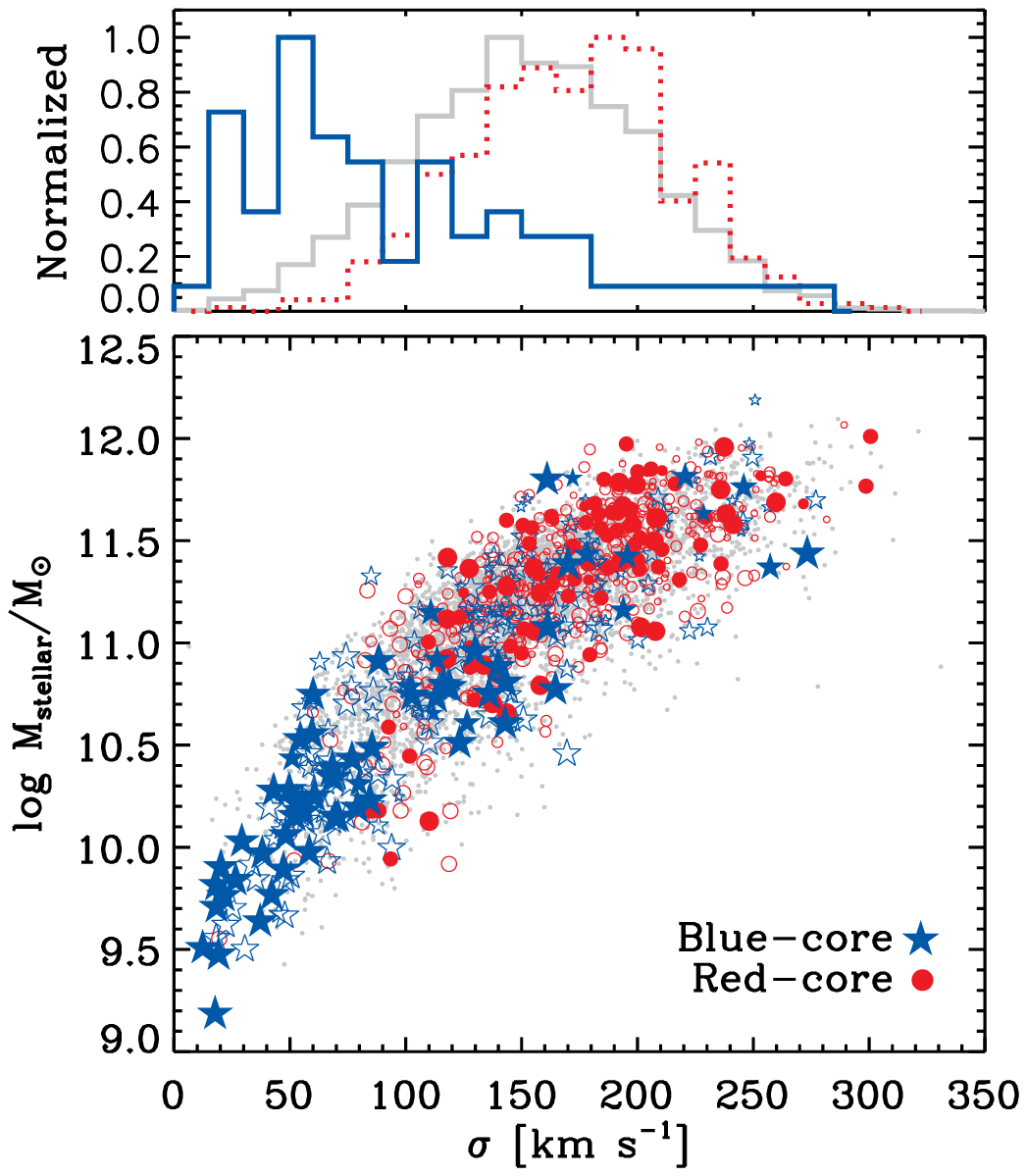}
\caption{{\it Top}: The distribution of velocity dispersion for our
early-type galaxies. The gray histogram represents the total sample.
Blue solid and red dotted histograms represent blue-cored and red-cored
galaxies, respectively. {\it Bottom}: Stellar mass of early-type galaxies
versus velocity dispersion. Blue-cored galaxies tend to
have lower stellar velocity dispersion, while red-cored
galaxies are more massive. The symbols are the same as in
Figure~\ref{fig:err_slope}.} \label{fig:Slope_vel}
\end{figure}

The stellar velocity dispersion is correlated with a galaxy mass
\citep[e.g.][]{FJ76}.
Aperture corrections are applied to the observed velocity dispersion due to the fact
that the fixed-diameter fibers sample larger portions of galaxies at larger distance, systematically
underestimating the true velocity dispersion \citep{Bernardi02,Ka03,Tre04,Ke05,Ke08}.
The applied aperture corrections follow \citet{Cap06}, namely:
\begin{equation}
\rm \sigma_{\rm corr} = (\frac{R_{\rm fiber}}{R_{e}})^{0.066 \pm 0.035} \sigma_{\rm fiber},
\end{equation}
where R$_{\rm fiber}$ is 1.5 arcsec and R$_{\rm e}$ is the effective radius.

We show the distribution of velocity dispersions for our sample of early-type
galaxies in the top panel of Figure~\ref{fig:Slope_vel}. We find 
that red-cored galaxies (red dotted) are relatively massive compared 
to average early-type galaxies (gray) while blue-cored galaxies (blue solid)
tend to have lower stellar velocity dispersion, once more in 
agreement with studies at moderate redshift \citep[e.g.][]{Ferreras09,Kan09}.
In the bottom panel of Figure~\ref{fig:Slope_vel}, we also plot the 
velocity dispersion versus stellar mass.
The stellar masses were derived by fitting the $u, g, r, i$ and $z$ SDSS photometry
using a two-burst scenario \citep[e.g.][]{fs00, Kaviraj07, S07b}
with the stellar models of \citet{Maraston05}. Variable parameters include the age and
mass-fraction of the young burst, the age of the old burst and the amount of
dust extinction following the \citet{Calzetti00} law.
We measure stellar masses by minimizing the $\chi^{2}$ statistic.
It is interesting to notice that the transition between blue- and red-cored
galaxies occurs at around $M_s\sim 0.5-1\times 10^{10}$M$_\odot$, which
is the threshold below which the stellar populations of elliptical
galaxies become younger \citep{Rogers08}.

%%%%%%%%%%%%%%%%%%%%%%%%%%%%%%%%%%%%%%%%%%%%
\subsection{H$\beta$ absorption line}

% figure 6
\begin{figure}
\centering
\includegraphics[width=0.5\textwidth]{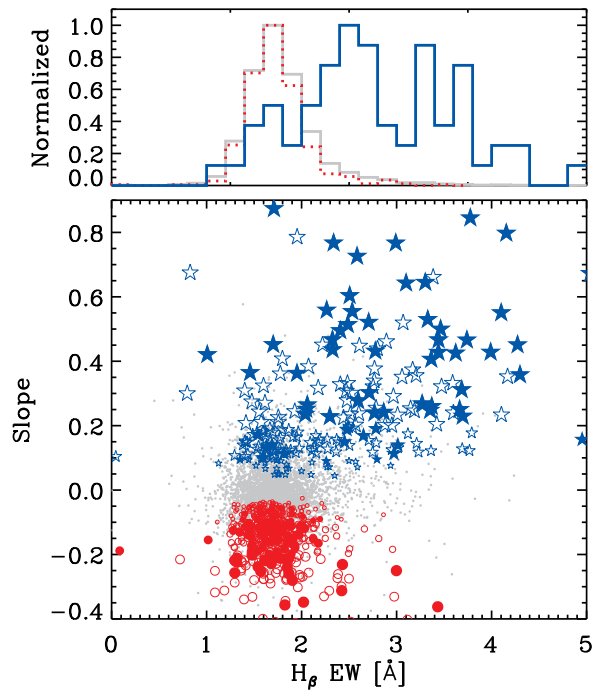}
\caption{{\it Top}: The distribution of H$\beta$ absorption line
strength for our early-type galaxies. The gray histogram represents
the total sample, blue solid histogram represents blue-cored galaxies and red dotted
histogram represents red-cored galaxies. {\it Bottom}: $g-r$ radial slope of
early-type galaxies versus H$\beta$ absorption line
strength. Blue-cored galaxies show stronger H$\beta$ absorption
than red-cored galaxies. The symbols are the same as in
Figure~\ref{fig:err_slope}.} \label{fig:Slope_Hb}
\end{figure}

The presence of strong Balmer absorption lines betray the presence of
main sequence stars formed in the past $\sim 1$~Gyr.  In
Figure~\ref{fig:Slope_Hb}, we show in the top panel the distribution
of the equivalent width (EW) of H$\beta$ for our early-type
galaxies. The distribution of H$\beta$ EWs for red-cored galaxies
(red dotted) is similar to that of the whole sample (gray). 
Interestingly, blue-cored galaxies (blue solid) have a distribution of
H$\beta$ strengths that peaks at a roughly 1.5 \AA\ higher EW than the
general sample.  This is more evidence towards recent star formation
in blue-cored galaxies. In the bottom panel, we show a trend
between the H$\beta$ EW and the $g-r$ color gradient. We note that blue-cored
galaxies with steeper slopes tend to show enhanced H$\beta$ line
strengths, which would imply that young stellar populations in the central
region are responsible for the color gradients.

%table3
\begin{deluxetable*}{llrrrrr}
\tablewidth{0pt}
\tablecaption{Spectral line Classification Results}
\tablehead{
\colhead{Classification} & 
\colhead{} & 
\colhead{Total} &  
\colhead{strong blue-cored$^{1}$} & 
\colhead{blue-cored} & 
\colhead{red-cored}
}
\startdata
Number  &   & 5002 (100\%)   & 66 (100\%)    & 216 (100\%) & 545 (100\%) \\
Strong emission$^{2}$  &  & 549 (10.97\%)   & 44 (66.67\%)  & 85 (39.35\%) & 60 (11.01\%) \\
 & Starforming... &  185 (3.70\%)    & 29 (43.94\%) & 47 (21.76\%) &   6 (1.01\%) \\
 & Transition...... &  176 (3.52\%)    & 5 (7.58\%)     & 12 (5.55\%)   &  19 (3.48\%) \\
 & Seyfert........... &   80 (1.60\%)     &  8 (12.12\%)  & 15 (6.94\%)   &  3 (0.55\%) \\
 & LINER........... &   107 (2.14\%)   & 2 (3.03\%)    & 11 (5.09\%)   &  32 (5.87\%) \\
Strong H$\beta$ absorption$^{3}$  &   & 1020 (20.39\%) & 53 (80.30\%) & 127 (58.79\%) & 79 (14.49\%) \\
Quiescent$^{4}$  &    & 3836 (76.69\%) & 7 (10.61\%) & 75 (34.72\%) & 421 (77.25\%) \\
\enddata
\tablenotetext{1}{Blue-core galaxies which have steeper slopes than 1$\sigma$ confidence levels (filled stars).} 
\tablenotetext{2}{Measurement of all four lines has S/N greater than 3.}
\tablenotetext{3}{H$\beta$ EW greater than 2 \AA\ .}
\tablenotetext{4}{Neither measurement of all four lines has S/N greater than 3 nor H$\beta$ EW greater than 2 \AA\ .}
\label{tbl:bpt_class}
\end{deluxetable*}

%%%%%%%%%%%%%%%%%%%%%%%%%%%%%%%%%%%%%%%%%%%%
\subsection{Emission line diagnostics}

% figure 7
\begin{figure}
\centering
\includegraphics[width=0.5\textwidth]{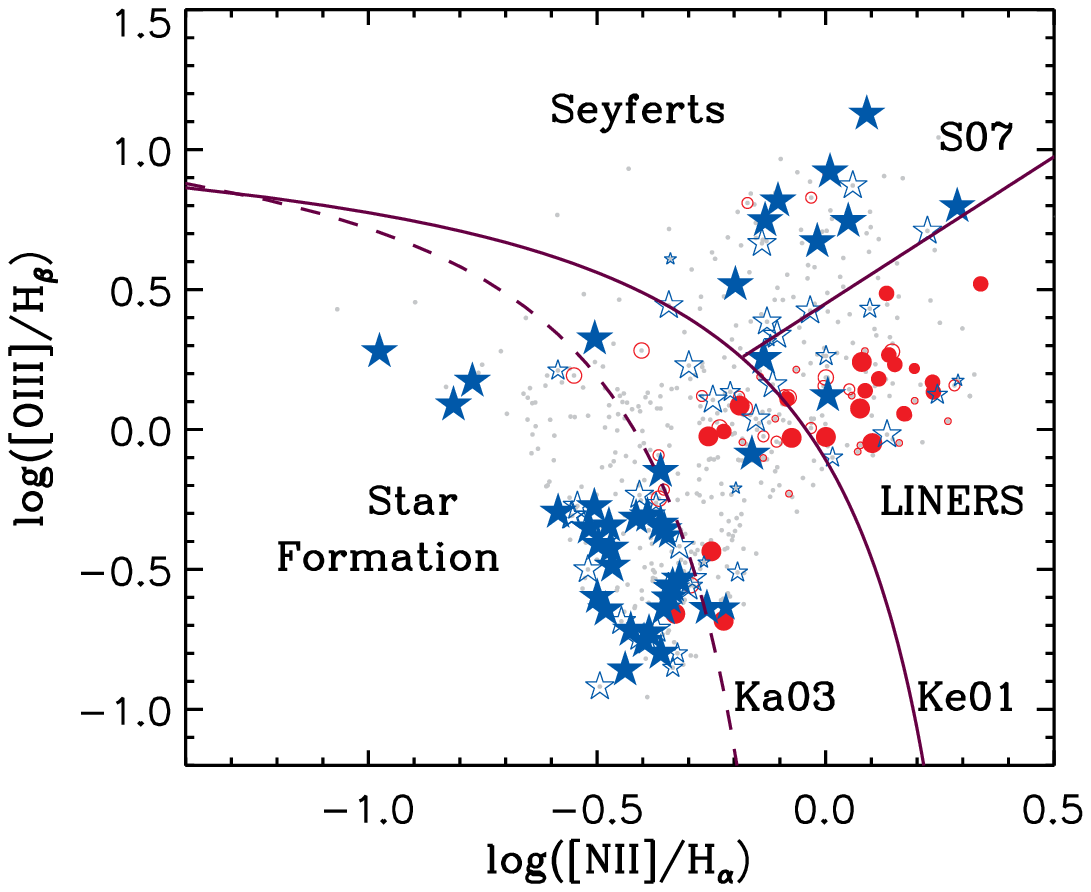}
\caption{The emission line diagnostic diagram for our sample
\citep{BPT1981}. We only plot galaxies where [\ion{N}{2}],
H$\alpha$, [\ion{O}{3}] and H$\beta$ lines are detected with S/N
$>$ 3. The curve labeled Ka03 is the empirical purely star-forming llimit of
\citet{Ka03}, while the curve labeled Ke01 represents the theoretical maximum
starburst model from \citep{Ke01}. All galaxies between the Ka03 and
the Ke01 are transition systems that they do have emission lines with contributions
from both star-formation and AGN. The line from S07
\citep{S07b} divides Seyferts and LINERS.  The symbols are the same as
Figure~\ref{fig:err_slope}.} \label{fig:BPT}
\end{figure}

%figure 8
\begin{figure}
\centering
\includegraphics[width=0.5\textwidth]{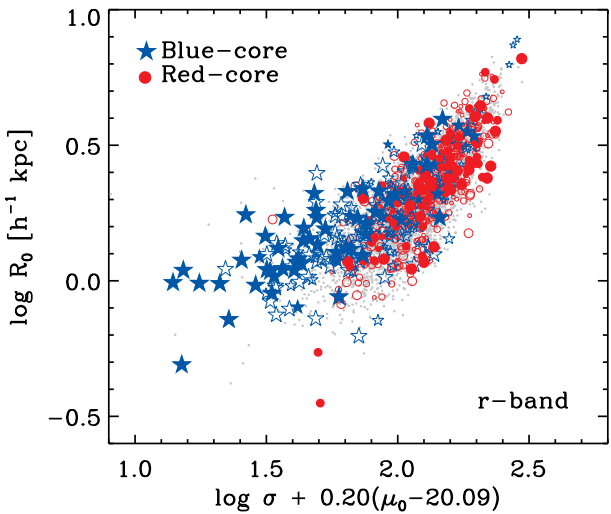}
\caption{The Fundamental Plane of early-type galaxies. blue-cored
and red-cored galaxies are marked with blue stars and red circles,
respectively. Gray points in the background indicate all early-type
galaxies. There are indications that red- and blue-cored galaxies may lie on different planes.
The symbols are the same as in
Figure~\ref{fig:err_slope}.} \label{fig:FP_r}
\end{figure}

% figure 9
\begin{figure*}
\centering
\includegraphics[width=1\textwidth]{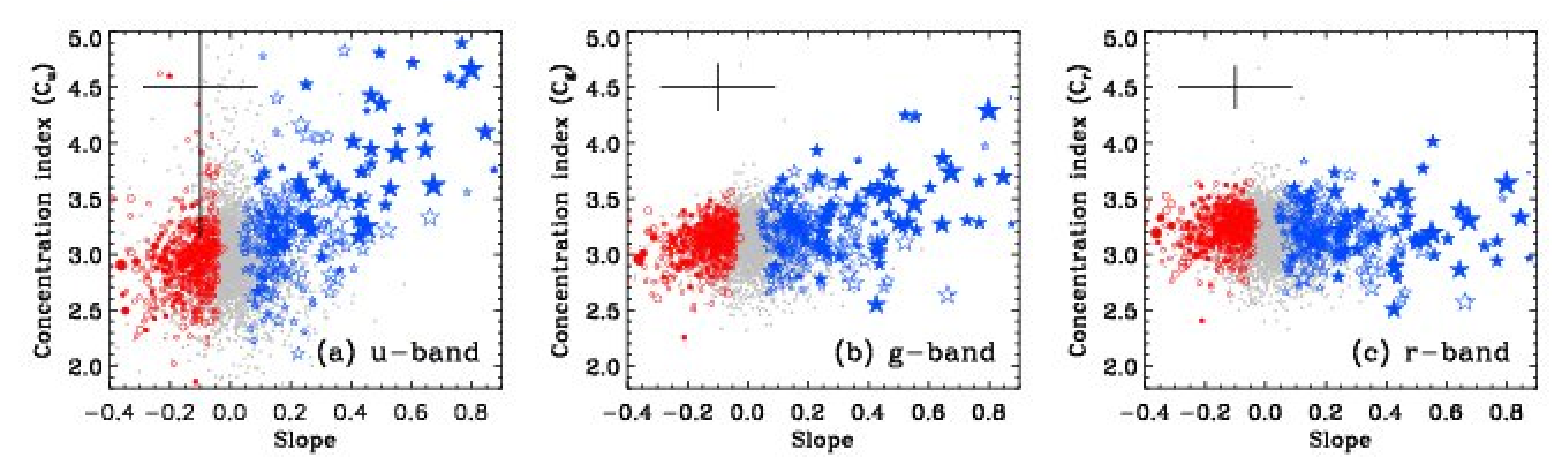}
\caption{The {\it u, g} and {\it r} band concentration indices of
early-type galaxies as a function of $g-r$ slope. Gray dots in the
background indicate the total sample of visually classified early-type
galaxies. The symbols are the same as Figure~\ref{fig:err_slope}, but the
symbol size represents H$\beta$ absorption line strength. The typical 1-sigma
error in the measurements is shown in each panel. The slope errors are measured
in this study, and the concentration index errors are adopted from the SDSS
database.}
\label{fig:C_slope}
\end{figure*}

Emission-line diagnostic diagrams are useful tools to distinguish
between star formation and AGN activity. The BPT diagram
\citep{BPT1981} allows the classification of galaxies with (narrow)
emission lines into star-forming galaxies and AGN. AGN are also split
into Seyfert II AGN and LINERs (Low Ionization Nuclear Emission Line
Region). In Figure~\ref{fig:BPT}, we plot the BPT diagram using the
emission line ratios [\ion{O}{3}]/H$\beta$ and [\ion{N}{2}]/H$\alpha$.
Only galaxies with a signal-to-noise (S/N) above 3 \citep{Ka03} in all
four lines are shown in the diagram. We use the demarcation line by \citet{Ka03}
(dashed curve) to separate star-forming galaxies. The remaining
galaxies are divided into AGN and transition region. The latter
correspond to a blend of star-forming and AGN galaxies, and are
located between the empirical line of \citet{Ka03} and the theoretical
maximum starburst model of \citet{Ke01} (solid curve). AGN are
subclassified into Seyferts and LINERs as shown by the straight solid line
\citep{S07b}.

In Table~\ref{tbl:bpt_class}, we show the results of the spectral line
classification. Most early-type galaxies are virtually free of
emission lines ($\sim$89 per cent; that is, total(100) minus strong emission(10.97)).
Twenty three per cent of total early-type 
galaxies (that is, total(100) minus quiescent(76.69)) show significant emission 
lines (S/N $>$ 3 in all four BPT lines) or strong H$\beta$ absorption 
lines ($EW(H\beta) > 2$).  
We find that about 65~per cent (that is, total(100) minus quiescent(34.72)) of blue-cored galaxies show
strong H$\beta$ absorption and/or emission lines. 
Most of the emission line ratios are consistent with ongoing star
formation. Only 10~per cent (quiescent; 10.61 per cent) of blue-cored galaxies with 
slopes steeper than the 1$\sigma$ confidence level (i.e. strong positive color gradient
galaxies) show neither emission nor strong Balmer absorption lines.  
Interestingly, some blue-cored galaxies are classified as Seyferts.  Therefore,
we note that blue-cored galaxies are mainly due to centrally
concentrated star formation and a smaller fraction could be associated 
with Seyfert activity. Most red-cored galaxies, on the other hand, are
classified as quiescent (78.98 per cent).  If red cores
show emission lines, most of their emission line ratios are consistent
with LINER AGN. red-cored galaxies appear $not$ to be host to
active star formation or powerful AGN.
\\

%%%%%%%%%%%%%%%%%%%%%%%%%%%%%%%%%%%%%%%%%%%%
\subsection{Fundamental Plane}

Early-type galaxies exhibit a scaling relation between the effective
radius $R_{\rm e}$, the effective mean surface brightness $\mu_{e}$ and
central velocity dispersion $\sigma$ (i.e. the Fundamental Plane,
FP, \citealt{detal87,dd87}). We compare here the location of blue-cored and red-cored galaxies on the
FP. In order to construct the fundamental plane of our early-type
galaxies, we have to apply a number of corrections as follows.  We convert an
elliptical aperture to an effective circular radius \citep{Bernardi03}
and apply a k-correction and the cosmological dimming effect:
\begin{equation}
\rm R_{e} \equiv \sqrt{b/a} R_{dev},
\end{equation}
\begin{equation}
\rm \mu = m_{dev} + 2.5 log(2\pi (R_{e})^{2}) - K(z) - 10 log(z+1),
\end{equation}
where b/a is the ratio of the minor and major axes, $R_{\rm dev}$ and
$m_{\rm dev}$ are the effective radius and  apparent magnitude,
respectively, from the de Vaucouleurs fit \citep{deVau48}.

We show in Figure~\ref{fig:FP_r} the FP of the total
sample (gray), blue-cored galaxies (blue stars) and red-cored
galaxies (red circles).
Red- and blue-cored galaxies appear to lie on different planes. A more detailed
analysis is on the way.
We will discuss this discrepancy via stellar population modelling in \S 4.2.

%%%%%%%%%%%%%%%%%%%%%%%%%%%%%%%%%%%%%%%%%%%%
\subsection{Concentration index}

The concentration index is often used as a quantitative measure of
galaxy morphology \citep[see e.g.][]{Morgan57, Conselice06} and is a useful tool
because the surface brightness distribution of a galaxy depends strongly on
its formation history.  The concentration
index, C, is defined as the ratio between R$_{90}$ and R$_{50}$ from the SDSS pipeline
parameters, which are the radii that encompass 90 and 50 per cent, respectively, of the
Petrosian flux, namely:
\begin{equation}
\rm C_{x} = \frac{R_{x90}}{R_{x50}},
\end{equation}
where x ={\it u, g, r\ } labels the SDSS band used to determine the
concentration \citep{Stoughton02}.
After unresolved sources (which just result in a PSF),
early-type galaxies are the most concentrated systems with
C$_{r} > 2.6$ \citep[see Figure~1 of ][]{Ferreras05}.

In Figure~\ref{fig:C_slope}, we show the {\it u, g} and {\it r} band
concentration indices versus the slope of $g-r$ color for
our sample of early-type galaxies. 
The color code for the symbol is the same as in Figure 2.
The size of the symbols is
proportional to the H$\beta$ line strength.  The right panel
indicates that blue-cored and red-cored galaxies show similar concentration
index in the {\it r} band. However, the other panels show that a significant
fraction of blue-cored galaxies are more concentrated in {\it u} and {\it g} bands,
while there is a tendency for red-cored galaxies to be less concentrated, although
examples of galaxies showing the opposite behavior can also be identified. 
We note that the most concentrated blue-cored galaxies in {\it u} band
tend to also show strong H$\beta$ absorption lines, indicating large fractions
of young populations in the center. This difference as well as the increase in scatter
are naturally explained by the different sensitivity of the passbands to young stars.

%%%%%%%%%%%%%%%%%%%%%%%%%%%%%%%%%%%%%%%%%%%%
\subsection{Environment}

% figure 10
\begin{figure}
\centering
\includegraphics[width=0.5\textwidth]{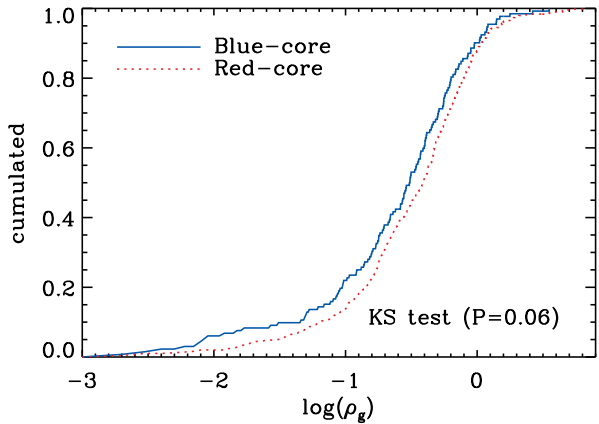}
\caption{Cumulative distribution of the local density
log($\rho_{g}$). In order to minimize a selection bias with stellar mass,
this figure shows both blue-cored and red-cored
galaxies in the same stellar mass bin (10.5 $<$ log(M/M$_{\bigodot}$)
$<$ 12). The KS test discriminates between these two distributions
at a 95\%
confidence level. blue-cored galaxies tend to prefer slightly lower density regions
compared to red-cored galaxies.} \label{fig:density}
\end{figure}

We find evidence that a large fraction of blue-core galaxies
have undergone recent star formation from H$\beta$ absorption line
strengths and emission line diagnostics. We now address the possible
cause of this recent star formation in early-type galaxies. One
possible scenario involves galaxy mergers and interactions.  If
galaxies experience merging events involving gas-rich late-type
galaxies, it might result in enhanced star formation during the
encounter. We investigate this possibility by looking at
environment and also by checking the nature and location of the
nearest neighbor.

% figure 11
\begin{figure}
\centering
\includegraphics[width=0.5\textwidth]{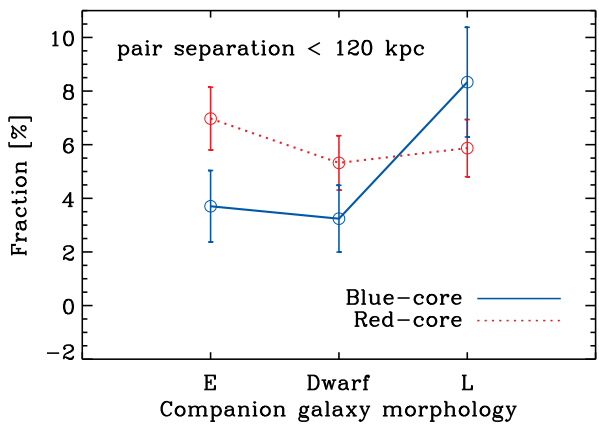}
\caption{Fraction of pair galaxies with respect to (visually classified)
morphology. The morphologies of pair galaxies are divided into
three types: early-type galaxy (E), dwarf galaxy (Dwarf) and
late-type galaxy (L). Blue-cored galaxies appear more are likely to have a
late-type companion.} \label{fig:pair_type}
\end{figure}

% figure 12
\begin{figure}
\centering
\includegraphics[width=0.5\textwidth]{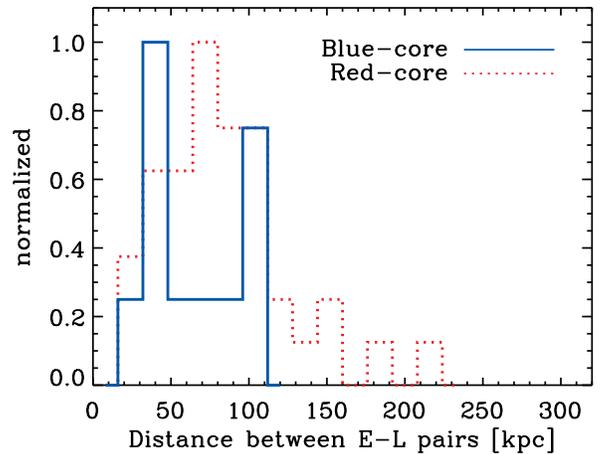}
\caption{The distribution of separation in close pairs (within 300~kpc)
comprising an early- and a late-type galaxy. Pairs with a blue-cored
galaxy are found at shorter separations with respect to 
pairs with a red-cored galaxy.}\label{fig:pair_hist}
\end{figure}

Figure~\ref{fig:density} shows the cumulative distribution of the
local density parameter $\rho_{g}$ defined by \citet{S07a} \citep[see
also][]{Yoon08}.  This density parameter ($\rho_{g}$) represents a
weighted local number density, and is obtained by counting all
neighbors within 2~Mpc with a Gaussian-weighting scheme. In order to
avoid selection bias with respect to mass, we limit both blue-cored
and red-cored galaxies to the same stellar mass
range between 10.5 $<$ log(M/M$_{\bigodot}$) $<$ 12. We perform the
Kolmogorov-Smirnov test (KS test) and find that blue-cored and
red-cored galaxies are not consistent with being drawn from the
same parent population at the 95 per cent confidence level.
These two cumulative distributions of
$\rho_{g}$ show that blue-cored galaxies prefer lower density
environments compared to red-cored galaxies.
However, this may also come from the fact that more massive galaxies
reside in denser regions. For a narrower mass bin, we find less clear
density dependence. But we had to use such as large mass range for this
statistical test because the number of blue-cored galaxies in our sample is limited.
We would need a much larger sample to break the mass-density degeneracy.

We also consider the nearest neighbor galaxies and their
morphology. Two galaxies are considered to be in a close pair if 
r$_{p}$ $<$ 120 kpc (where $r_{\rm p}$ means projected physical separation) 
and $\Delta$z $<$ 0.001. 
We show the fraction of galaxies in close pairs in Table~\ref{tbl:pair_frac}.  
Radial color gradients seem insensitive to pair galaxy fraction.
However, this analysis is hampered by projection effects that can
mimic true pairs which are more serious in denser regions.

If the recent star formation in the central region of blue-cored
galaxies is caused by gas inflows from a neighboring galaxy, one might
wonder whether the morphology of a companion galaxy 
-- which correlates with gas fraction -- could affect color gradients.
To test this point, we perform visual classification of the companion galaxies. The
morphology of the companion galaxies are classified into three types:
early-type (E), dwarf (Dwarf) and late-type (L). 
Figure~\ref{fig:pair_type} shows the fraction of pair galaxies
with respect to the companion morphology. Taken at face value, blue-cored
galaxies have more late-type companions than red-cored
galaxies. 

Figure~\ref{fig:pair_hist} shows the distribution of pair separation
in systems comprising an early- and a late-type galaxy. We find that
pairs with a blue-cored galaxy have smaller separations than
those with a red-cored galaxy, in agreement with the picture that a
nearby, gas-rich late-type galaxy can contribute to the onset of
residual star formation. However, we should emphasize here that most
(92 per cent) of the blue-cored galaxies in our sample do not have a
late-type companion. The presence of a late-type companion may
affect the star formation of the galaxy in question, but not all
star-forming early types are caused by the presence of a late-type companion. 
Hence, other mechanisms (such as minor mergers with nearby gas
rich dwarf galaxies or even gas clouds) should be invoked to explain
the general trend.

%table4
\begin{table}
 \centering
  \caption{Pair galaxy fraction}
  \begin{tabular}{@{}l r@{}}
  \tableline \tableline
   Classification & Fraction (r$_{p} <$ 120 kpc) \\
   \tableline
   Total          & 764/5002 (15.27\%) \\
   blue-cored  &  45/262 (17.18\%) \\
   red-cored  & 120/647 (18.54\%) \\
   \tableline \tableline
   \end{tabular}
\label{tbl:pair_frac}
\end{table}

%figure 13
\begin{figure}
\centering
\includegraphics[width=0.5\textwidth]{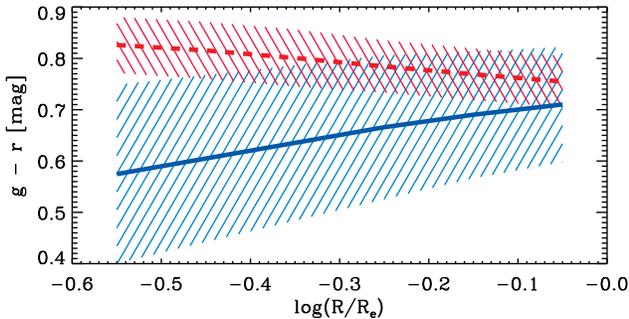}
\caption{The mean radial $g-r$ color gradients of blue-cored
(blue solid) and red-cored (red dotted) galaxies. The shading corresponds to
the scatter within the sample, given at a 1$\sigma$ confidence level.}
\label{fig:Obs_meangr}
\end{figure}

%figure 14
\begin{figure}
\centering
\includegraphics[width=0.5\textwidth]{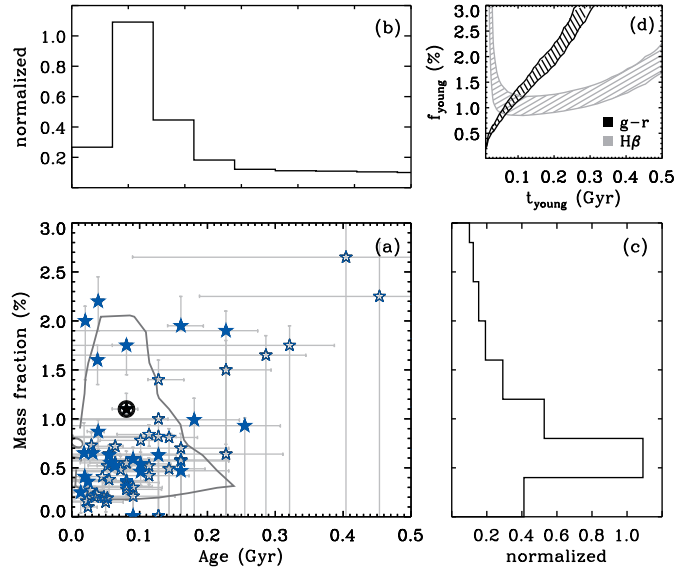}
\caption{The normalized likelihood distribution in age
and mass fraction for the young stellar component.
(a) The star symbols represent the best fits to the blue-cored galaxies
(as classified in Figure 2) along with error bars in grey.
Only those with $\chi_{\rm red}^2 <3$ are shown here.
The contour gives the 1$\sigma$ confidence level
from a Monte Carlo simulation (see text for details).
(b) Normalized age distribution. (c) Normalized mass fraction of the
young stellar component. (d) A typical $\chi^2$ marginalization
for a galaxy. This particular case is for the circled galaxy in (a).
} \label{fig:mcs_chi}
\end{figure}

%%%%%%%%%%%%%%%%%%%%%%%%%%%%%%%%%%%%%%%%%%%%
\section[]{Discussion}
\label{sec:model}

%figure 15
\begin{figure*}
\centering
\includegraphics[width=0.9\textwidth]{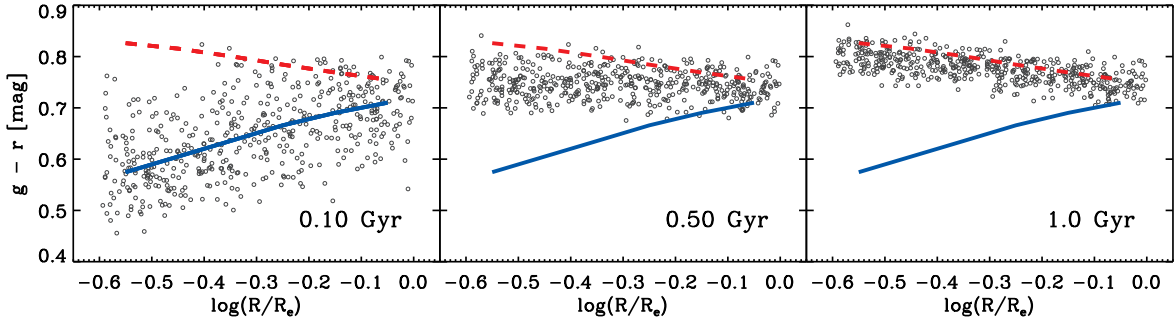}
\caption{The evolution of the color gradient for blue-cored galaxies.
The blue solid and red dotted lines represent the mean radial $g-r$ color gradients
of blue-cored and red-cored galaxies, respectively. The open circles correspond to
models using the observations of red-cored and blue-cored galaxies as
constraints on a two-burst scenario (see text for details).  As the
populations age (from left to right), the blue-cored galaxy transforms into a
red-cored galaxy.} \label{fig:evol_sim}
\end{figure*}

%figure 16
\begin{figure}
\centering
\includegraphics[width=0.5\textwidth]{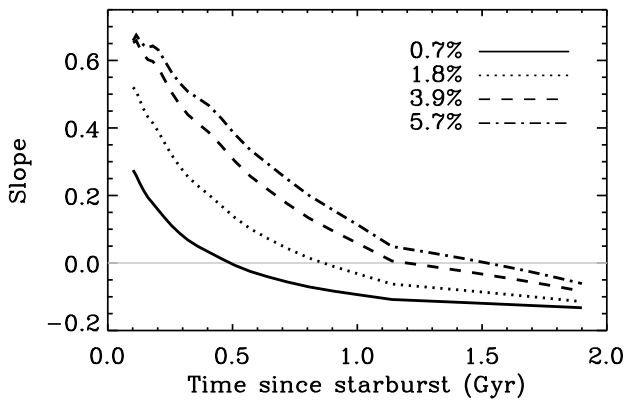}
\caption{Evolution of slope obtained from the probability distribution
in the age and mass fraction of the young component. We show how the
slope evolves for various possible choices of the young stellar mass
fraction. We choose mass fractions from the peak of the probability
distribution (solid), or at the 1$\sigma$ (dotted), 2$\sigma$ (dashed)
and 3$\sigma$ (dot-dashed) confidence levels. The slope of blue-cored 
galaxies decreases rapidly from positive to negative in about 1.5 Gyr
after the star formation episode.}
\label{fig:evol_color}
\end{figure}

%%%%%%%%%%%%%%%%%%%%%%%%%%%%%%%%%%%%%%%%%%%%
\subsection{Age and mass fraction of young stars}

%figure 17
\begin{figure*}
\centering
\includegraphics[width=0.9\textwidth]{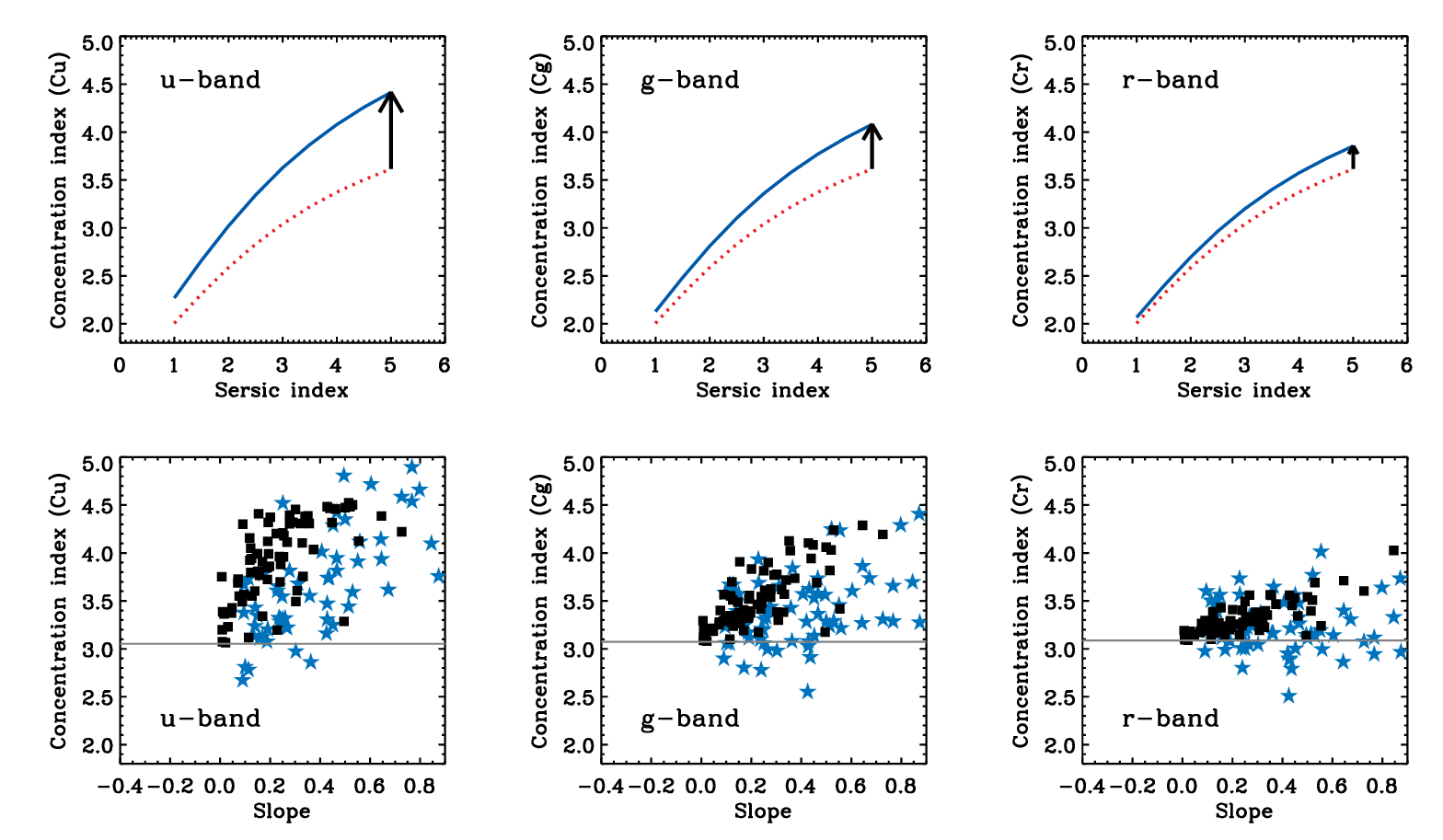}
\caption{{\it Top}: Concentration indicies of {\it u, g} and {\it r}
band as a function of S$\acute{e}$rsic index. The red dotted line represents
the concentration index assuming a purely old population.  The blue solid
line shows the effect of a centrally concentrated young population on
each passband.  {\it Bottom}: Concentration indicies versus slope.
The stars represents observed
blue-cored galaxies and the squares show the results from the
modelling (see text for details). The gray horizontal line marks the 
mean concentration index of the total sample in each passband.}
\label{fig:evol_con}
\end{figure*}

The aim of this section is to quantify the amount of recent star
formation in our sample. In order to study the radial distribution of
young stars, we construct the mean $g-r$ radial color gradient of the
sample, shown in Figure~\ref{fig:Obs_meangr}. The dotted and
solid lines indicate red-cored and blue-cored galaxies, respectively.
The shading in each case represents the 1$\sigma$ confidence level.
Color gradients in early-type galaxies are usually a result of the underlying stellar
populations, mainly in age and metallicity. It is difficult, however, to
distinguish the effects of a small change in age from those of a
small change in metallicity. Most red-cored galaxies, however,
tend to show overall red $u-r$ colors (see \S 3.1) and weak H$\beta$
line strengths (see \S 3.3), suggesting an old population.
Therefore, we consider the negative slopes of red-cored galaxies as
a result of a simple metallicity gradient and assume that
blue-cored galaxies have the same metallicity gradient. 
This assumption is also motivated by the
lack of evolution in the color gradient of red-cored
galaxies with redshift \citep{Ferreras05,Ferreras09}.

To calculate the characteristic metallicity of each region in
the galaxy, we assume a single uniform age of 12~Gyr but change
the metallicity distribution until it matches the observed $g-r$ color.
The base model in this study are those of \citet{Yi03} which are
specialized for old populations. Since those models
do not cover ages younger than 1~Gyr, however, we combine them with the models
of \citet{BC03}. According to the models, the central region in 
red-cored galaxies has a metallicity distribution with a
peak around the solar value, while at the effective radius the peak
decreases to half the solar.

We consider a two-stage star formation history to determine the age
and mass fraction of the young stellar component of blue-cored
galaxies. This model combines an old and a young stellar population
\citep{fs00, Kaviraj07, S07b}. The old population has a fixed age of
12~Gyr with a metallicity gradient determined from the mean radial
color gradient of red-cored galaxies. The young component has the solar
metallicity and is allowed to vary in age ($0.01 \le t_{young} \le
$10~Gyr) and mass fraction ($10^{-4} \le f_{young} \le 1$). To
constrain these two parameters, we fit the observed $g-r$ color and
H$\beta$ EW to models and compute the associated $\chi^{2}$ statistic
to obtain a probability distribution of the age and mass fraction of
the young stellar population. 
We further perform a Monte Carlo simulation with $10^3$ realizations
to explore the parameter space for minima and the associated errors
around them.
The best fit and confidence levels are shown in 
Figure~\ref{fig:mcs_chi}(a) and also in the accompanying histograms
(b and c). In panel (d) we show the $\chi^2$ minimization performed
on a sample galaxy, circled star in (a). These clearly show that young 
($\sim 10^2$ Myr old) stellar populations are present in the central region
and the stellar mass fraction of the young component reaches up to
a couple of per cent.

By evolving the young component in time, we can derive the evolution
of the slope. Figure~\ref{fig:evol_sim} shows the evolution of model blue-cored
galaxies, assuming the mean properties and scatters for the young 
population derived here.
The color gradients
observed suggest that this young component should have a decreasing
contribution with increasing radius. Hence, we choose a central value
for the mass fraction in young stars of 0.7~per cent and decrease this
mass fraction outward in a manner that reproduces the observed $g-r$
profiles. 
Its mean age is assumed to be 100~Myr to begin with, and we monitor
the color gradient as time goes on. The solid and dashed lines
represent the mean radial $g-r$ color gradients of blue-cored and red-cored
galaxies, respectively. The color gradients caused by the
young population revert to the fiducial values of red-cored galaxies
after about 1~Gyr, as expected given the small fractions in young
stars. 
In Figure~\ref{fig:evol_color} we show the evolution of the change in slope
with respect to different values of the mass fraction in young stars.
We choose four different cases based on the probability distribution
of the modeling (see Figure~\ref{fig:mcs_chi}). We consider the
peak of the probability distribution (0.7 per cent, solid), and the 1$\sigma$
(1.8 per cent, dotted), 2$\sigma$ (3.9 per cent, dashed), and 
3$\sigma$ (5.7 per cent, dot-dashed) confidence levels.
The slope of the blue-cored galaxies changes rapidly
from positive to negative.  The positive gradients in early-type galaxies
must be transient features that are visible only for $\sim$0.5 -- $\sim$1.3 billion years
or so after a centrally-concentrated star formation episode
\citep[see figure~8 of ][]{Ferreras09}.

%figure 18
\begin{figure*}
\centering
\includegraphics[width=0.8\textwidth]{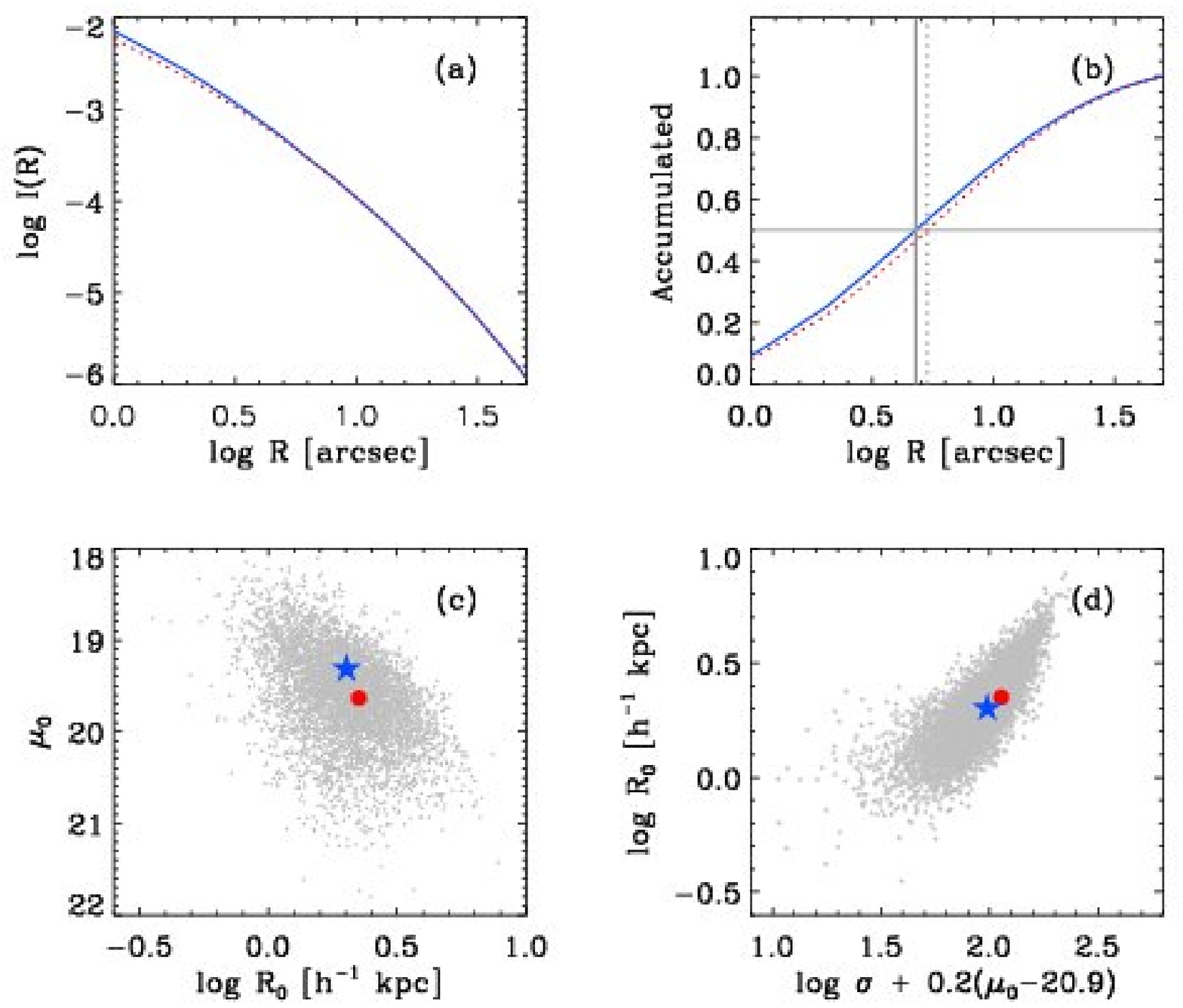}
\caption{(a) The surface brightness distribution. The dotted line
represents the intensity profile assuming a purely old population. The
intensity profile is normalized to I(0). The solid line shows the
combined light profile of the old and young populations. 
(b) Cumulative light profile.  Dotted and solid lines represent the underlying
old population and the combined model, respectively. The vertical
lines represent the effective radius of each profile. 
(c) The effect of a centrally concentrated young population on the Kormendy
relation. Red circle and blue star represent old and total (old plus
young) model, respectively. The gray dots in the background indicate
all observed early-type galaxies. 
(d) The effect of a centrally-concentrated young population on the FP. 
Symbols are the same as in (c).}
\label{fig:FP_modeling}
\end{figure*}

% figure 19
\begin{figure*}
\centering
\includegraphics[width=0.8\textwidth]{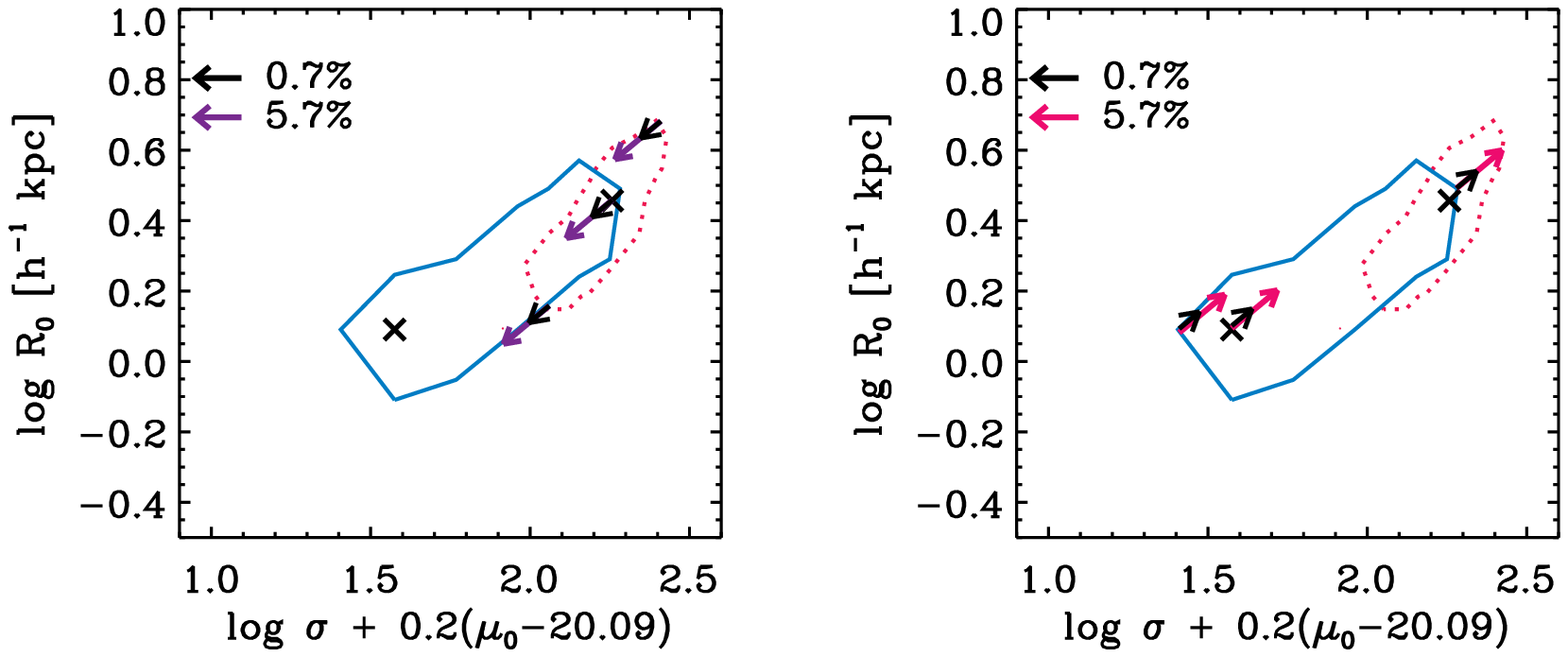}
\caption{The effect of a centrally concentrated episode of star
formation on the FP. Blue solid and red dotted lines delimit the
observed FP (shown at the 1$\sigma$ confidence level)
for blue-cored and red-cored galaxies, respectively. The vectors show the
contribution of a 100~Myr young components with a mass fraction of
0.7\% or 5.7\%. {\it Left}: Effect of a
young component on red-cored galaxies.  {\it Right}: Effect of ageing
of the young component of blue-cored galaxies.}
%In both cases we find
%that blue/red-cored galaxies belong to different distributions,
%separated by velocity dispersion (or stellar mass). Only the most
%massive blue cores can be explained by the presence of a significant
%($\sim$6\%) amount of centrally concentrated, recent star formation.}
\label{fig:FP_contour}
\end{figure*}

%%%%%%%%%%%%%%%%%%%%%%%%%%%%%%%%%%%%%%%%%%%%
\subsection{The effect of young stars}

We now discuss the effect of the young stellar populations on the
concentration index and the FP.  To explore the effect
of young stars on these properties, we compute the change of the surface
brightness distribution of a galaxy due to recent star formation in the
central region.  We start from the S$\acute{e}$rsic profile:
\begin{equation}
I(R) = I(0)\exp[-k(R/R_{e})^{1/n}]
\end{equation}
For $n \ge 1$, $k$ satisfies the relation $k = 2n - 0.324$ \citep{Ciotti91}.

We assume that the intensity profile of the underlying old population
has the metallicity gradient constrained by the color gradient and
that the young stellar populations are more centrally concentrated
to satisfy the observed positive color gradients. 
A model effective radius of 5~arcsec is chosen for this exercise.  
We construct the total light profile as the superposition of
the surface brightness profile for the old and young populations. This
profile allows us to quantify the effect of a young component on the
effective radius, mean surface brightness or concentration index.

%%%%%%%%%%%%%%%%%%%%%%%%%%%%%%%%%%%%%%%%%%%%
\subsubsection{Concentration index}

The concentration index is a useful tool because the surface brightness
distribution of a galaxy depends strongly on its formation history.
Star formation, mergers and accretion events might
affect the surface brightness distribution by rearranging
the stellar components in the progenitors as well as by forming new
stars in geographically biased ways. 
For example, if star formation is triggered towards the center, the
concentration index will increase.

Figure~\ref{fig:evol_con} shows the concentration index of the old
(dotted) and the mixed (solid) populations.  The top panels of
Figure~\ref{fig:evol_con} show that the strongest effect is on
the concentration measured in the {\it u} band because of its highest
sensitivity to young stars. We also show the increase of the
concentration index as a function of slope in the bottom panels. The
horizontal line indicates the mean concentration index for the
total sample in each passband.  Using the estimated best fit age and
mass fraction of the young stellar component of blue-cored
galaxies (see \S 4.1), we compare our models (black squares)
with observations (stars).  The concentration index in the model
shows a clear correlation with the radial color gradient slope.
The correlation is stronger for shorter wavelengths.
The match seems poor for the $u$ band. This is probably because of
the inaccuracy of our model assumptions. We assumed that recent
star formation occurred only in the central region, from center to effective radius.
However, the SDSS concentration indices are measured from the whole radial range.
If the residual star formation extends outside the effective radius,
it would lower our model concentration index to match the data better.
But such a detailed modeling is beyond the scope of this investigation.
Despite the simplicity of the model considered here, we successfully
match the observed concentration indices in various bandpasses.

%%%%%%%%%%%%%%%%%%%%%%%%%%%%%%%%%%%%%%%%%%%%
\subsubsection{Fundamental plane}

In order to examine the effect of the young population on the
FP, we calculate the change of the effective radius and
the mean surface brightness due to recent star formation in the
central region. We assume that the surface brightness distribution of a
red-cored early-type galaxy follows a de~Vaucouleurs' profile (i.e. Sersic
index $n=$ 4).

In Figure~\ref{fig:FP_modeling}(a), we show the
intensity profile of the underlying old population (dotted) and the
combined profile (solid).  We show the cumulative luminosity
profiles in Figure~\ref{fig:FP_modeling}(b). 
A population with young stars at the center
tends to have a smaller effective radius and a brighter mean surface 
brightness, as shown for the Kormendy relation in
Figure~\ref{fig:FP_modeling}(c). In order to study the effect of this
centrally concentrated young population on the FP, we
compare our models with observations, as shown in
Figure~\ref{fig:FP_modeling}(d). This suggests that although the
models predict a significant offset in the Kormendy relation, the
effect is more subtle on the FP. 
Hence, the clear difference between the {\sl observed} blue- and red-cored
galaxies on the FP
(Figure~\ref{fig:FP_r}) seems beyond what can be modeled by a simple addition of
young stars. Instead, one needs to consider that blue-cored galaxies indeed have on
average lower velocity dispersions than the general sample.

To further illustrate this point, Figure~\ref{fig:FP_contour} shows
the effect of centrally concentrated recent star formation on the
FP. The solid and dotted contour lines delimit the region 
occupied by the FP for blue-cored and red-cored
galaxies, respectively. The black and purple/pink arrows
correspond to the presence of 0.7 or 5.7 per cent of a 100~Myr population
at the center. The left panel considers what would happen if this
young component is added to the population of red cores. On the right
panel, we consider the fading of this young component in blue-cored galaxies,
as the galaxies age.
In both cases we find that blue/red-cored galaxies belong to different distributions,
separated by velocity dispersion (or stellar mass). 
One can clearly see that although the most 
massive blue-cored galaxies could be explained by a recent
episode of star formation (comprising 6~per cent of its stellar mass
content, which is hardly unjustifiable), 
for the majority of (low mass) blue-cored early-type galaxies,
one must consider that they correspond to systems
with markedly lower velocity dispersion than red-cored galaxies.

%%%%%%%%%%%%%%%%%%%%%%%%%%%%%%%%%%%%%%%%%%%%
\section{Summary}
\label{sec:summary}

We present radial $g-r$ color gradients of visually classified
early-type galaxies at $0.00 \le z \le 0.06$ selected from Data
Release 6 of the Sloan Digital Sky Survey. Due to the shallow exposure
of DR6 color gradients are difficult to measure in the bulk of the
sample. The brightest galaxies exhibit a negative color
gradient (centrally redder) as expected for their observed radial
metallicity variations. However, a significant fraction -- roughly
30~per cent which might depend on data selection strategy --
features blue cores (positive color
gradients). Evidence is presented, suggesting that these blue-cored
galaxies have undergone a recent episode of star formation that is
centrally concentrated. Most are blue also in terms of integrated
$u-r$ colors and show strong central H$\beta$ absorption line
strengths and/or emission ratios that are consistent with star forming
populations.  Combining SDSS and {\it GALEX} UV photometry, we find that all
blue-cored galaxies show blue UV$-$optical colors. They also
tend to have a lower stellar velocity dispersion.  We investigate their
environmental dependence and star formation properties. Blue-cored
galaxies exhibit a tendency to live in lower density environments
compared to red-cored galaxies and, if found in close pairs, have a
preference for a late-type companion.  On the other hand,
red-cored galaxies, which are relatively massive, are
located on the red sequence as shown in the integrated $u-r$
color-magnitude relation, and those with emission lines reside in the
LINER region of a BPT diagram.  

Based on a
simple model that overlays a young stellar component over an old
population, our sample gives a mass fraction in young stars below
2~per cent with age 100$-$800~Myr in the central regions of blue-cored galaxies.
Furthermore, these positive color gradients in
early-type galaxies are visible only for $\sim$0.5 -- $\sim$1.3 billion years after a
star formation event.  Although this centrally located star formation
episode can decrease the effective radius and increase the mean
surface brightness, the FP of the (majority of) low
mass blue-cored galaxies cannot be reproduced by any amount of recent
star formation on a red-cored galaxy. Instead, they require a lower
velocity dispersion, i.e. blue-cored and red-cored early-type galaxies
are, perhaps fundamentally, different  populations, mainly split with respect to stellar
mass or velocity dispersion.

From the perspective of our investigation on the optical color
gradients, we conclude that star-forming early types are not 
simply the star-forming counterparts of the quiescent ones.
While many parameters are connected to each other in degenerate manners, 
the most important role seems to be played by galaxy mass.
This {\em star-formation} dichotomy (between the star-forming and quiescent 
early types) is consistent with that of \citet{S06nature} and also  
with the {\em kinematic} dichotomy previously addressed
\citep[e.g.][]{Davies83, Bender88, KB96, Faber97, Rest01, Lauer07, Kimm07}.
Recent theoretical models \citep[e.g.][]{Khochfar09} explain
the {\em star-forming} dichotomy in connection with the {\em kinematic}
dichotomy through distinct merging histories by mass.
More massive early types are interpreted as results of equal-mass
dry mergers. This is compatible to the axis ratio distribution analysis
of \citet{Kimm07}. 
On the other hand, small early types are suggested to be a product of
early-type and late-type merger, in which case the cold gas from the
late-type component would linger in the merger remnant and
get used for residual star formation. \citet{Kaviraj09} further
point out that it was more likely minor mergers than major ones that 
was involved in the formation of low-mass early types exhibiting
signs of residual star formation. 
All these works together seem to suggest a coherent picture for the 
formation of early-type galaxies.

\section*{Acknowledgments}
Numerous comments from the anonymous referee helped us improve the
quality of the manuscript significantly.
We thank Sugata Kaviraj, Marc Sarzi, Young-Wook Lee, Changbom Park, 
Taysun Kimm, and Yun-Young Choi for useful discussions.
This study would not have been possible without the wealth of
publicly available data from the SDSS. 
This work was supported by the Korea Science and Engineering
Foundation (KOSEF) grant to SKY funded by the Korea government
(MEST) (No. 20090078756).
IF was supported by a grant from the Royal Society.
Support for the work of KS was provided by NASA through Einstein
Postdoctoral Fellowship grant number PF9-00069 issued by the 
Chandra X-ray Observatory Center, which is operated by the 
Smithsonian Astrophysical Observatory for and on behalf of NASA 
under contract NAS8-03060. KS gratefully acknowledges support 
from Yale University.
This project made use of the {\it GALEX} ultraviolet data, the HyperLeda database
and the NASA/IPAC Extragalactic Database.

\clearpage

\end{document}